\documentclass[prd,aps,preprintnumbers,floats,floatfix,superscriptaddress,preprintnumbers,
showpacs,eqsecnum,nofootinbib,twocolumn]{revtex4-2}
\usepackage[utf8]{inputenc}
\usepackage{latexsym,array,theorem,mathrsfs,bm,float}
\usepackage{psfrag}
\usepackage{amsfonts,amsmath,amssymb,latexsym,array,afterpage,
theorem,mathrsfs,bm,float,epsfig,color,graphicx,tabularx,here,multirow}
\usepackage[linktocpage=true]{hyperref}

\newcommand{\nn}{\nonumber \\}
\newcommand{\Mpl}{M_{\rm Pl}}
\newcommand{\D}{{\rm d}}
\newcommand{\Lcal}{\mathcal{L}}

\newcommand{\Fcal}{\mathcal{F}}
\newcommand{\Vcal}{\mathcal{V}}

\begin{document}
\baselineskip=12pt

\preprint{YITP-22-67}
\title{Resolving the pathologies of self-interacting Proca fields: \\
A case study of Proca stars
}
\author{Katsuki Aoki}
\affiliation{
Center for Gravitational Physics and Quantum Information, Yukawa Institute for Theoretical Physics, Kyoto University, 606-8502, Kyoto, Japan
}

\author{Masato Minamitsuji}
\affiliation{Centro de Astrof\'{\i}sica e Gravita\c c\~ao  - CENTRA, Departamento de F\'{\i}sica, Instituto Superior T\'ecnico - IST, Universidade de Lisboa - UL, Av. Rovisco Pais 1, 1049-001 Lisboa, Portugal}

\date{\today}

\begin{abstract}
It has been argued that a self-interacting massive vector field is pathological due to a dynamical formation of a singular effective metric of the vector field, which is the onset of a gradient or ghost instability. We discuss that this singularity formation is not necessarily a fundamental problem but a breakdown of the effective field theory (EFT) description of the massive vector field. By using a model of ultraviolet (UV) completion of the massive vector field, we demonstrate that a Proca star, a self-gravitating condensate of the vector field, continues to exist even after the EFT suffers from a gradient instability without any pathology at UV, in which the EFT description is still valid and the gradient instability in EFT may be interpreted as a standard dynamical instability of a high-density boson star from the UV perspective. On the other hand, we find that the EFT description is broken before the ghost instability appears. This suggests that a heavy degree of freedom may be spontaneously excited to cure the pathology of the EFT as the EFT dynamically tends to approach the onset of the ghost instability.

\end{abstract}

\maketitle

\section{Introduction}

On the one hand, although all the existing observational data on astrophysical and cosmological scales
are consistent with the predictions of general relativity \cite{Clifton:2011jh,Will:2014kxa},
current and future probes of strong gravity regions
including gravitational wave astronomy 
provide new opportunities to test general relativity \cite{Berti:2018cxi,Berti:2018vdi,Barack:2018yly}.
On the other hand, 
while the latest data of gravitational waves are all consistent with 
the emissions from black hole and neutron star mergers in general relativity,
hypothetical (horizonless) compact objects based on various theoretical grounds
have been extensively proposed \cite{Cardoso:2017cqb,Cardoso:2019rvt},
which will be tested with various observational channels \cite{Berti:2018cxi,Berti:2018vdi,Barack:2018yly},
depending on the compactness and nature of objects.

Boson stars
that are gravitationally bound nontopological solitonic objects \cite{Jetzer:1991jr,Schunck:2003kk,Visinelli:2021uve,Liebling:2012fv}
are known as the representative exotic horizonless compact objects.
A complex scalar field constituting a boson star
has the oscillatory time dependence $e^{-i\omega t}$ with the frequency $\omega$.
Boson stars are characterized by 
the two conserved charges, 
the Arnowitt-Deser-Misner (ADM) mass $M$ and  
the Noether charge $Q$ associated with the global $U(1)$ symmetry.
Since $M$ and $Q$ correspond to the gravitational mass and the scalar particle number,
respectively,
boson stars are gravitationally bound if $\mu Q>M$ where $\mu$ is the mass of the complex scalar field.
Boson star solutions were first constructed 
in the massive complex scalar theory $\mu^2 |\phi|^2/2$ \cite{Kaup:1968zz,Ruffini:1969qy,Friedberg:1986tp}.
It is known that
the properties of boson stars 
are characterized in the $M-\omega$ (and $Q-\omega$) relations.
In the limit of the vanishing scalar amplitude,
$M\to 0$ and $Q\to 0$ while $\omega\to \mu$,
the spacetime approaches the Minkowski spacetime.
As one increases the scalar amplitude at the center,
while $\omega$ decreases,
$M$ and $Q$ first increase, 
reach their maximal values at the same $\omega$,
and then decrease until $\omega$ reaches the minimum value,
which forms the first branch of the boson star solutions,
where $\mu Q>M$ and boson stars are gravitationally bound.
The solutions with smaller central amplitude before reaching the maximum of $M$ and $Q$
are also dynamically stale 
\cite{Gleiser:1988rq,Gleiser:1988ih,Hawley:2000dt}.
After reaching its minimal value,
$\omega$ increases and decreases repetitively and eventually converges to a single value
as the central amplitude increases,
forming the second, third, and higher branches in the $M (Q)-\omega$ relations,
where always $M>\mu Q$ and boson stars are dynamically unstable.
In the massive complex scalar field theory,
the maximal mass of boson stars is of ${\cal O} (\Mpl^2/\mu)$,
where $\Mpl$ is the (reduced) Planck mass,
while in the presence of the quartic order self-interaction $\lambda |\phi|^4$
this is of ${\cal O} (\sqrt{\lambda}\Mpl^3/\mu^2)$ 
which can be much higher than the pure massive case for $\mu\ll \Mpl$
\cite{Colpi:1986ye}.
Boson stars in the presence of other self-interacting scalar potentials
have been extensively studied \cite{Schunck:2003kk,Guerra:2019srj}.

Boson star solutions can be naturally extended to other bosonic fields,
especially the complex vector (Proca) field, which are known as Proca stars
\cite{Brito:2015pxa,Brihaye:2017inn,Garcia:2016ldc,Minamitsuji:2018kof,Herdeiro:2020jzx,Cardoso:2021ehg,Zhang:2021xxa,Jain:2022nqu} (see also \cite{Aoki:2017ixz,Brito:2020lup,Jain:2021pnk} for spin-2 solitonic objects).
In the massive complex Proca theory $\mu^2 \bar{A}^\mu A_\mu/2$,
where $A_\mu$ is the complex vector field and $\bar{A}_\mu$ is the complex conjugate of $A_\mu$, 
Proca star solutions exist for an arbitrary large amplitude of the Proca field at the center of the star,
and  their properties are very similar to
 those of  scalar boson stars \cite{Brito:2015pxa}.
As in the case of scalar boson stars,
Proca star solutions can be divided into different branches in the $M(Q)-\omega$ relations,
and 
only the first branch solutions which have the smooth limit to the Minkowski solution for the vanishing Proca amplitude
and the maximal values of $M$ and $Q$,
are energetically stable, i.e., $\mu Q>M$.
Due to the healthy properties of the massive Proca theory,
Proca star solutions of this type have been extensively applied to various astrophysical problems
\cite{Sanchis-Gual:2018oui,DiGiovanni:2020ror,Bustillo:2020syj,Herdeiro:2021lwl,Rosa:2022tfv,Rosa:2022toh}.

The situation, however, is drastically changed 
when the self-interaction potential of the complex Proca field $V(\bar{A}^\mu A_{\mu})$,
such as the quartic-order self-interaction
$\lambda (\bar{A}^\mu A_{\mu})^2/4$,
is taken into consideration besides the Proca mass term $\mu^2 \bar{A}^\mu A_\mu/2$.
In Ref. \cite{Minamitsuji:2018kof},
in the presence of 
the quartic-order self-interaction
as well as the mass term, $V=\mu^2 \bar{A}^\mu A_\mu/2+\lambda (\bar{A}^\mu A_{\mu})^2/4$,
it has been shown that
irrespective of the sign of $\lambda$
Proca star solutions
cease to exist
for the central Proca amplitude at a critical point
whose value depends on the other parameters in the model.
In the presence of  the sextic-order self-interaction as well, 
similar results were obtained in \cite{Herdeiro:2020jzx,Cardoso:2021ehg}.
Proca star solutions as well as scalar boson star solutions
are constructed numerically
by integrating the field equations 
under the regularity conditions at the center
and the exponential decay of the field at the spatial infinity.
The problem arises when the first-order radial derivative of the radial component 
of the Proca field diverges at a certain radius,
beyond which one cannot integrate the field equations numerically.
As we see later, 
at the singular point the radial component of the effective metric 
for the self-interacting Proca field vanishes
and 
the perturbations lose the hyperbolicity.
Thus, the problem may be interpreted as the onset of the 
so-called gradient instability (see below) at the background level.
Such a problem is absent for the Proca field without a self-interaction,
while could generically arise for a self-interacting Proca field.

Recently, 
a conceptually related problem was pointed out in \cite{Clough:2022ygm,Coates:2022qia,Mou:2022hqb},
which claim that
a self-interacting Proca field suffers from a generic ghost instability.
Note that 
the `ghost' and `gradient' instabilities are associated with the wrong signs of the kinetic and gradient terms
in the Lagrangian for the perturbations on a given background,
respectively.
Unlike the tachyonic instability, 
i.e.,~the instability associated with the wrong sign of the mass term,
which appears only for long wavelength modes and grows with a finite rate,
the ghost and gradient instabilities
grow arbitrarily fast 
{\it if they continue existing in arbitrarily high-energy/momentum scales.}
The perturbation theory breaks down within an infinitesimally short time scale, and no prediction is trustable.
In the context of the time domain analysis, 
the onset of the ghost or gradient instability
could be interpreted as the breakdown of the hyperbolic evolution of the perturbations
and hence the well-posedness of the initial value problems.
Ref.~\cite{Clough:2022ygm}
performed
a numerical simulation of the late-time evolution of the superradiant growth 
of the self-interacting  (real) Proca field on the Kerr background,
and 
showed that
when the self-interaction becomes important
irrespective of the sign of the coupling constant
the time derivative of the temporal component of the Proca field
diverges at a certain moment of time,
beyond which one cannot follow the time evolution.
At this moment of time
the temporal component of the effective metric for the Proca field 
vanishes,
and the self-interacting Proca field suffers from a ghost instability.
The problem of a ghost instability 
is expected to be generic to the self-interacting Proca sector,
and independent of the background spacetime geometry.
Ref.~\cite{Coates:2022qia} studied
the propagation of the self-interacting Proca wave
in $1+1$ dimensional Minkowski spacetime,
and
showed that it inevitably suffers from a ghost (or gradient) instability.
Once the role of space and time is reversed,
the situation is very similar to 
the gradient instability problem for Proca stars with a self-interaction mentioned above.


In the case that the self-interacting Proca field is a fundamental field,
the ghost or gradient instability is indeed the pathology of the theory.
On the other hand,
in the case that the self-interacting Proca theory 
is a low-energy effective description of a more fundamental theory,
the problem may be avoided 
once ultraviolet (UV)  physics,
for instance the dynamics of heavy fields,
is properly taken into consideration.
The aim of this work is to suggest a solution to this problem
from the effective field theory (EFT) viewpoint,
and propose a (partial) UV completion to the self-interacting Proca theory,
where 
the onset of the gradient instability 
simply indicates a breakdown of the self-interacting Proca theory as an EFT.
Although we will not address a dynamical simulation of the Proca field,
our method should be directly applied and can provide insight into the ghost problem during the time evolution.

The paper is organized:
In Secs.~\ref{sec:proca} and \ref{sec:procastar},
we review the propagation of 
perturbations around a nontrivial background configuration of the self-interacting complex Proca field,
and
Proca star solutions in the presence of the quartic-order self-interaction,
respectively.
In Sec.~\ref{sec:partialUV},
we introduce a model of the partial-UV completion of 
the self-interacting Proca theory.
In Sec.~\ref{sec:uvprocastar},
we study the Proca star solution 
in the partial-UV completion theory of the quartic-order self-interaction
and demonstrate
how the inclusion of a heavy field $\chi$ maintains the hyperbolicity of the equations of motion even at the singularities of the effective metric.
The last section \ref{sec:summary}
is devoted to giving a brief summary and conclusion.

\section{Propagations of self-interacting complex Proca field}
\label{sec:proca}

We consider a self-interacting complex massive (spin-1) Proca field $A_{\mu}$ 
minimally coupled to gravity, 
which is  described by the action
\begin{align}
S_{\rm Proca}= \int \D^4 x \sqrt{-g}\left[ \frac{\Mpl^2}{2}R
-\frac{1}{4e^2}F_{\mu\nu}\bar{F}^{\mu\nu} -V(X) \right]
\label{action:Proca}
\end{align}
where 
$R$ is the Ricci scalar associated with the metric $g_{\mu\nu}$,
$g:={\rm det}(g_{\mu\nu})$,
$\Mpl^2=1/8\pi G$ is the Planck mass with $G$ being the gravitational constant,
\begin{align}
F_{\mu\nu}:=2\partial_{[\mu}A_{\nu]}\,, \quad X :=A_{\mu}\bar{A}^{\mu}
\end{align}
and the bar means the complex conjugate. 
$V(X)$ represents the self-interacting potential of the complex Proca field.
The parameter 
$e$ can be set to unity by normalizing the Proca field, but we keep it for later convenience. 
Since the complex Proca field enjoys a global $U(1)$ symmetry, there exists the associated Noether current
\begin{align}
j^{\mu}:=\frac{i}{2e^2}\left( \bar{F}^{\mu\nu}A_{\nu} - F^{\mu\nu}\bar{A}_{\nu} \right)
\,.
\end{align}
The Proca field $A_{\mu}$ may be suffered from a pathology (e.g.,~ghost instability) 
in the presence of nonlinear self-interaction, 
which is most easily understood by taking the decoupling limit. 
Let us replace the Proca field $A_{\mu}$ according to
\begin{align}
A_{\mu} \to e A_{\mu} + \partial_{\mu} \phi
\end{align}
with $\phi$ a (complex) St\"{u}eckelberg field. We take the limit $e\to 0$ while keeping the potential finite. Then, the action \eqref{action:Proca} is reduced to a complex Maxwell field and a decoupled complex k-essence:
\begin{align}
S_{\rm Proca} \to \int \D^4 x \sqrt{-g}\left[ \frac{\Mpl^2}{2}R- \frac{1}{4} F_{\mu\nu}\bar{F}^{\mu\nu} -V(X_{\phi}) \right]
\,, 
\end{align}
where $X_{\phi} := \nabla_{\mu}\phi \nabla^{\mu} \bar{\phi} $.
The St\"{u}eckelberg field has a nonlinear kinetic term, suggesting that the kinetic term is not necessarily positive-definite around a nontrivial background. 

Let us elaborate on propagations of the Proca field. The equation of motion of the Proca field is
\begin{align}
\nabla^{\nu}F_{\nu\mu}-2e^2 A_{\mu} V' =0
\,, \label{eom:Proca}
\end{align}
where $V'$ may be regarded as the effective mass squared of the vector field. The prime represents the derivative with respect to the argument. The divergence of the equation of motion leads to a constraint equation
\begin{align}
\nabla^{\mu}(A_{\mu} V') = \nabla^{\mu}A_{\mu} V' + A_{\mu}\nabla^{\mu}V' = 0
\,. \label{diveom:Proca}
\end{align}

We study a high-frequency limit of the perturbations around a nontrivial background configuration of the Proca field. 
We continue to use $A_{\mu}$ to denote the background Proca configuration,
while we denote the perturbations of the Proca field by $\delta A_{\mu} + \partial_{\mu} \pi$ 
with the perturbations of the St\"{u}eckelberg field $\pi$. 
In the high-frequency limit, we only retain the highest derivative terms of the perturbations, so the equations \eqref{eom:Proca} and \eqref{diveom:Proca} yield
\begin{align}
\partial^2 \delta A_{\mu} - \partial^{\nu} \partial_{\mu} \delta A_{\nu} + \cdots  &=0 \,, \\
V' \partial^2 \pi + V'' ( A^{\mu} \bar{A}^{\nu} \partial_{\mu} \partial_{\nu} \pi + A^{\mu} A^{\nu} \partial_{\mu} \partial_{\nu} \bar{\pi} ) +\cdots &=0
\,, \label{diveom:Procahigh}
\end{align}
where $\cdots $ are terms that are at most linear in derivatives of the perturbations. 
Here, we may 
replace the covariant derivatives with the partial derivatives
since we are interested in the short wavelength limit
much smaller than the spacetime curvature scale.
The former equation is the Maxwell equation, representing the luminal propagations 
of the transverse modes of the Proca field,
and does not give rise to any pathology. On the other hand,
the propagations of the longitudinal modes may be modified due to the nonlinear interactions. By the use of the form $\pi= \pi^1 + i \pi^2$ where $\pi^1$ and $\pi^2$ are real fields, \eqref{diveom:Procahigh} can be rewritten as
\begin{widetext}
\begin{align}
\begin{pmatrix}
V'g^{\mu\nu}+2V'' A^{1\mu}A^{1\mu} & 2 V'' A^{1(\mu}A^{2\nu)} \\
2 V'' A^{1(\mu}A^{2\nu)} & V'g^{\mu\nu}+2V'' A^{2\mu}A^{2\mu}
\end{pmatrix}
\partial_{\mu}\partial_{\nu}
\begin{pmatrix}
\pi^1 \\ \pi^2
\end{pmatrix}
+\cdots
=0
\end{align}
where $A^1_{\mu}$ and $A^2_{\mu}$ are the real and imaginary parts of the background $A_{\mu}$. 
Moving to the Fourier space by $\pi^{1,2}=\int \D ^4k \, \pi^{1,2}_{(k)} e^{i k_\mu x^\mu}$
with $k_\mu$ being a four-wavevector,
the partial derivative may be replaced with $i k_{\mu}$ in the high-frequency limit, leading to the dispersion relations
\begin{align}
{\rm det}
\begin{pmatrix}
V'k^2+2V'' (A^1\! \cdot \! k)^2 & 2 V''  (A^1\! \cdot \! k) (A^2\! \cdot \! k) \\
2 V'' (A^1\! \cdot \! k) (A^2\! \cdot \!k) & V'k^2+2V''  (A^2\! \cdot \! k)^2
\end{pmatrix}
=V' k^2 g_{\rm eff}^{\mu\nu} k_{\mu} k_{\nu} = 0
\label{disrel}
\end{align}
\end{widetext}
with the effective metric
\begin{align}
g_{\rm eff}^{\mu\nu} := V' g^{\mu\nu} +  2V'' A^{(\mu} \bar{A}^{\nu)}
\,.
\label{effectivemetric}
\end{align}
The dispersion relation \eqref{disrel} implies that one of the longitudinal modes propagates on the spacetime metric $g^{\mu\nu}$ while the other propagates on the effective metric $g_{\rm eff}^{\mu\nu}$. The signature of the effective metric may differ from $(-,+,+,+)$ which signals the presence of an instability. The determinant of $g_{\rm eff}^{\mu\nu}$ is computed by
\begin{align}
{\rm det}(g_{\rm eff}^{\mu\nu}) &=  {\rm det}(g^{\mu\nu}) V'{}^2
\nn
&\times \left[ V'{}^2+ V''(  2X V' + V'' X^2 - V'' A_{\mu}A^{\mu} \bar{A}_{\nu} \bar{A}^{\nu}) \right]
.
\end{align}
Hence, the effective metric is singular at the point where
\begin{align}
V'=0
\,, \label{reg:effectivemetric1}
\end{align}
or
\begin{align}
V'{}^2+ V''(  2X V' + V'' X^2 - V'' A_{\mu}A^{\mu} \bar{A}_{\nu} \bar{A}^{\nu}) =0
\,,
\label{reg:effectivemetric2}
\end{align}
at which one of the longitudinal modes would be infinitely strongly coupled. The theory cannot be trusted when the effective metric $g^{\mu\nu}_{\rm eff}$ becomes singular.
Note that 
although each component of the effective metric \eqref{effectivemetric} is coordinate-dependent,
the physical conditions \eqref{reg:effectivemetric1} and \eqref{reg:effectivemetric2}
to determine the appearance of singular Proca star configurations
are expressed by scalar quantities and thus coordinate-independent.


\section{Proca star with quartic interaction}
\label{sec:procastar}

Before presenting a UV completion of the complex Proca field, let us discuss a concrete solution which tends to approach a singular effective metric.
As a concrete example, we use a quartic self-interaction
\begin{align}
V=\frac{\mu^2 }{2}A_{\mu}\bar{A}^{\mu} + \frac{\lambda}{4}(A_{\mu}\bar{A}^{\mu} )^2
\,,
\label{quarticpotential}
\end{align}
and study a relativistic self-gravitating condensate of the Proca field, known as the Proca star
\cite{Brito:2015pxa,Brihaye:2016pld,Garcia:2016ldc,Minamitsuji:2018kof,Herdeiro:2020jzx,Cardoso:2021ehg}, 
under the ansatz
\begin{align}
g_{\mu\nu}\D x^{\mu} \D x^{\nu} &= - \sigma^2(r)\left( 1- \frac{2m(r)}{r} \right) \D t^2
\nn
&+ \left( 1- \frac{2m(r)}{r} \right)^{-1} \D r^2 + r^2 \D \Omega_2^2
\,, \label{metricansatz} \\
A_{\mu}\D x^{\mu} &= e^{-i \hat{\omega} t }( a_0(r) \D t + i a_1(r) \D r)
\,, \label{Procaansatz}
\end{align}
where $a_0(r)$ and $a_1(r)$ are real functions of $r$ and $\hat{\omega}$ is a real and positive parameter. Note that although the vector field has both temporal and radial components with a time dependence $\propto e^{-i \hat{\omega} t }$, the effective metric \eqref{effectivemetric} is still static and diagonal under the above ansatz. Let us below explain the properties of the Proca star in the presence of the quartic interaction (see~\cite{Minamitsuji:2018kof} for details). Here, we set $e=1$ by normalizing the Proca field.

The basic equations are the Einstein equation and the equation of motion of the Proca field \eqref{eom:Proca}.
Note that the Proca field must satisfy the constraint equation \eqref{diveom:Proca} which reduces the order of differential equations. 
All the components of the Einstein equation are not independent thanks to the Bianchi identity; thus, we only need to use the $(tt)$ and $(rr)$ components as the independent equations. In addition, there are two independent equations from \eqref{eom:Proca} and \eqref{diveom:Proca} which, in combination with the Einstein equation, are enough to (numerically) solve the four functions $ \{ \sigma(r), m(r), a_0(r), a_1(r)\}$ under appropriate boundary conditions. 
By the use of \eqref{diveom:Proca} and the radial component of \eqref{eom:Proca}, we finally find the set of the first-order differential equations:
\begin{align}
\frac{\D \sigma(r) }{\D r} &= F_{\sigma}[\sigma, m, a_0, a_1,r] \,, \label{Peom:sigma} \\
\frac{\D m(r) }{\D r} &= F_m[\sigma, m, a_0, a_1,r]  \,, \\
\frac{\D a_0(r) }{\D r} &= F_0[\sigma, m, a_0, a_1,r]  \,, \\
\frac{\D a_1(r) }{\D r} &= F_1[\sigma, m, a_0, a_1,r]  \label{Peom:a1} \,, 
\end{align}
where $F_A~(A=\sigma,m,0,1)$ do not contain any derivatives. One can confirm that other components of the Einstein equation and \eqref{eom:Proca} are satisfied under \eqref{Peom:sigma}-\eqref{Peom:a1}.
Note that 
Ref.~\cite{Minamitsuji:2018kof} uses \eqref{diveom:Proca} and the radial component of \eqref{eom:Proca} to solve the equations of motion of the Proca field by which one obtains a second-order differential equation of $a_1$. Nonetheless, the same solutions are found by the first-order equation \eqref{Peom:a1} because the same boundary conditions are imposed.

A caveat is that $F_A$ may pass a singularity during the numerical integration. For instance, the functions $F_A$ diverge 
at a point where $r-2m=0$ which is a coordinate singularity of the present ansatz \eqref{metricansatz}. 
However, we will not encounter such a singularity 
since we consider a horizonless object. 
As pointed out in~\cite{Minamitsuji:2018kof}, $F_1$ is singular at
\begin{align}
g_{\rm eff}^{rr} = 0
\end{align}
which corresponds to a singularity of the effective metric \eqref{effectivemetric} and this singularity indeed appears in the solutions with the quartic interaction. Then, we cannot integrate the field equations numerically. The situation is analogous to what happens in the time evolution problem~\cite{Clough:2022ygm,Coates:2022qia,Mou:2022hqb}, by reversing the role of the time and the space.

Let us explicitly construct asymptotically flat solutions satisfying the following boundary conditions:
\begin{align}
m \to 0 \,, ~ \sigma \to \sigma_c \,, ~ a_0 \to a_c \,, ~ a_1 \to 0
\quad {\rm as} \quad r\to 0
\end{align}
and
\begin{align}
m \to m_{\infty}\,, ~ \sigma \to \sigma_{\infty}\,, ~ a_0 \to 0 \,, ~ a_1 \to 0
\quad {\rm as} \quad r\to \infty
\end{align}
where $\sigma_c, a_c, m_{\infty}$ and $\sigma_{\infty}$ are constant. Note that the ansatz \eqref{metricansatz} and \eqref{Procaansatz} has a freedom associated with a time rescaling $t\to ct, \omega \to c^{-1} \omega, \sigma \to c^{-1}\sigma, a_0 \to c^{-1} a_0$
with $c$ being a constant
which can be used to set, for example, $\sigma_c=1$ without loss of generality.
More precisely, the regularity condition at the center concludes that the solution 
must be given by the form:
\begin{align}
\label{solution_origin}
m(r)
&=
\frac{a_c^2}{24\Mpl^2 \sigma_c^4}
\left(
-3\lambda a_c^2
+2\mu^2 \sigma_c^2
\right)
r^3
+{\cal O} (r^4),
\nonumber\\
\sigma(r)
&=
\sigma_c
-
\frac{a_c^2}
        {4\Mpl^2\sigma_c^3}
\left(a_c^2\lambda-\mu^2\sigma_c^2 \right)
r^2
+{\cal O} (r^4),
\nonumber\\
a_0(r)
&=
a_c
-
\frac{a_c}{6\sigma_c^2}
\left(
a_c^2\lambda
-\mu^2\sigma_c^2
+{\hat \omega}^2
\right)
r^2
+
{\cal O} (r^4),
\nonumber\\
a_1(r)
&=
-\frac{a_c {\hat \omega}}
         {3\sigma_c^2 }r
+{\cal O} (r^3).
\end{align}
On the other hand, the asymptotic form of the solutions 
is given by 
$a_0, a_1\propto e^{-\sqrt{\mu^2-\omega^2}r}$
where $\omega $ is the proper frequency
for a distant observer defined by 
\begin{align}
\label{proper_freq}
\omega:=\frac{\hat \omega}{\sigma_\infty}.
\end{align}

For numerical computations, it is useful to use the following dimensionless variables
\begin{align}
\tilde{r}:=\mu r\,, ~ \tilde{m}:= \mu m\,, ~ \tilde{a}_0:=a_0/\Mpl\,, ~ \tilde{a}_1 := a_1/\Mpl
\label{normalizevar}
\end{align}
and the dimensionless parameters
\begin{align}
\tilde{\omega}:= \hat{\omega}/\mu\,, \quad \tilde{\lambda}:= \lambda \Mpl^2/\mu^2
\,.
\label{normalizepara}
\end{align}
Then, the parameters $\Mpl$ and $\mu$ do not appear in the equations of motion, so we do not need to fix them explicitly. Furthermore, we introduce
\begin{align}
\tilde{\alpha}_0 := \frac{\tilde{a}_0}{\sigma(1-2m/r)^{1/2} }\,, \quad
\tilde{\alpha}_1 := \tilde{a}_1 (1-2m/r)^{1/2}
\end{align}
by which the components of the (mixed) effective metric $\mathcal{H}^{\mu}{}_{\nu}:= 2g_{\rm eff}^{\mu\alpha} g_{\alpha\nu}/\mu^2$ are computed as
\begin{align}
\mathcal{H}^{\mu}{}_{\nu}&={\rm diag}[\mathcal{H}^t{}_t,~\mathcal{H}^r{}_r,~\mathcal{H}^{\theta}{}_{\theta},~\mathcal{H}^{\varphi}{}_{\varphi}]
\,, \\
\mathcal{H}^t{}_t&= 1+\tilde{\lambda}(-3\tilde{\alpha}_0^2+\tilde{\alpha}_1^2)
\,, \\
\mathcal{H}^r{}_r &= 1+\tilde{\lambda}(-\tilde{\alpha}_0^2+3\tilde{\alpha}_1^2)
\,, \\
\mathcal{H}^{\theta}{}_{\theta} &= \mathcal{H}^{\varphi}{}_{\varphi}  = 1+\tilde{\lambda}(-\tilde{\alpha}_0^2+\tilde{\alpha}_1^2) 
\,,
\end{align}
where $\theta$ and $\varphi$ are the angular coordinates. Note that $\mathcal{H}^t{}_t=0$ and $\mathcal{H}^r{}_r=0$ are the roots of \eqref{reg:effectivemetric2},
while $\mathcal{H}^{\theta}{}_{\theta}=\mathcal{H}^{\varphi}{}_{\varphi}=0$ is the root of \eqref{reg:effectivemetric1}. 
 Since \eqref{reg:effectivemetric2} is a scalar, $g^{rr}_{\rm eff} \propto \mathcal{H}^r{}_r =0$ is an actual (i.e.~not coordinate) singularity as long as $\mathcal{H}^t{}_t$ is finite.

\begin{figure}[t]
\centering
 \includegraphics[width=0.9\linewidth]{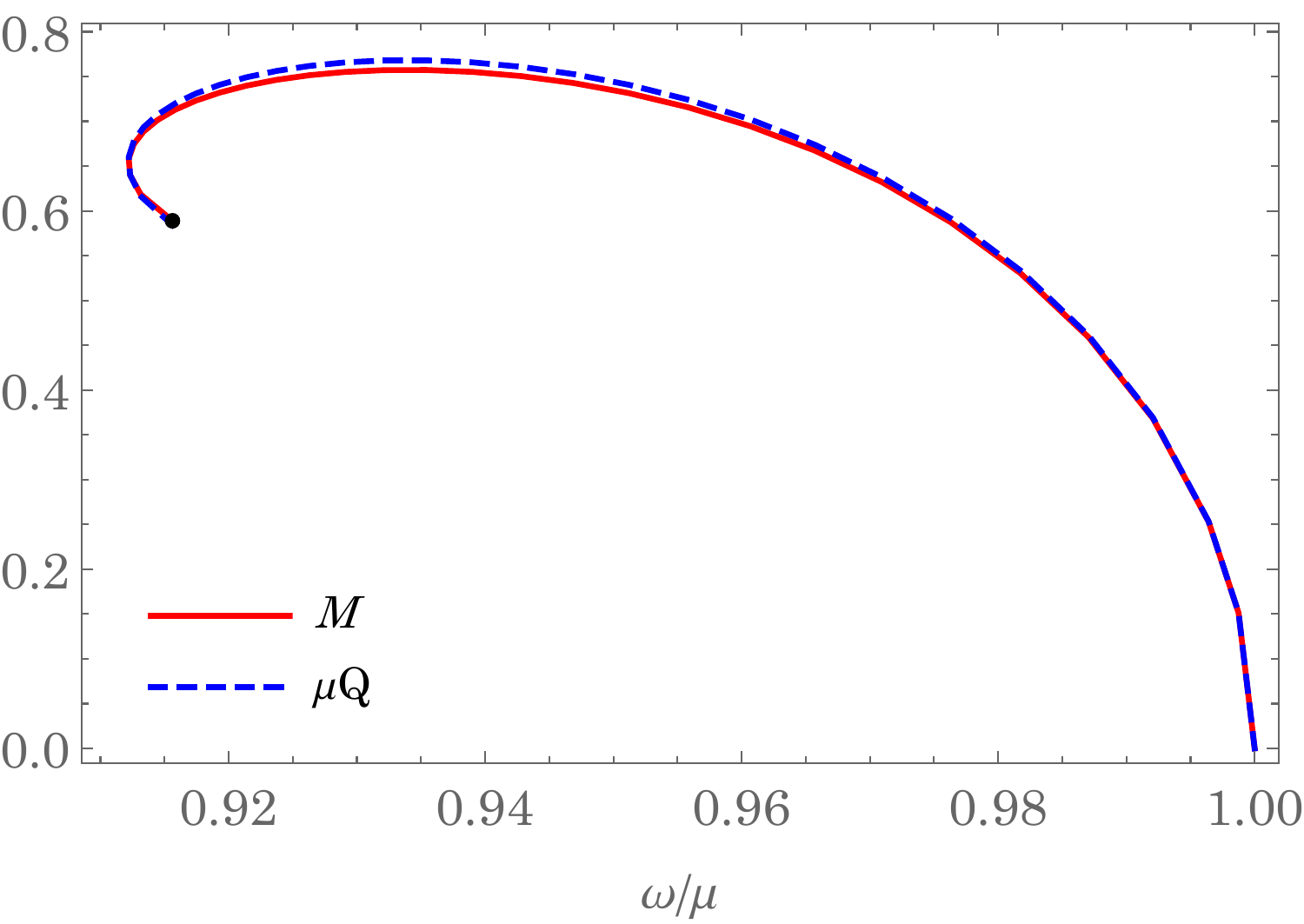}		
\caption{
$M-\omega$ (red-solid curve) and $Q-\omega$ (blue-dashed curve)
relations with $\tilde{\lambda}=-10$}
\label{figEFTomega}
\end{figure}

There exists a family of Proca stars and each solution is characterized by the ADM mass $M$, the Noether charge associated with the global $U(1)$ symmetry $Q$, and the eigenfrequency $\omega$ which are defined as
\begin{align}
M& :=  8\pi \Mpl^2 m_{\infty}\,, \\
Q& :=\int_{\Sigma} \D^3 x \sqrt{-g} j^t = \int^{\infty}_0 4\pi r^2 \D r \frac{a_1(\hat{\omega} a_1 - a_0')}{\sigma}
\,,
\end{align} 
respectively, where $\Sigma$ denotes the entire domain of a constant time hypersurface.
Fig.~\ref{figEFTomega}~represents $M$ and $Q$ as a function of $\omega/\mu$ for the quartic potential with $\tilde{\lambda}=-10$. The limit $\omega/\mu \to 1$ is the non-relativistic limit with the small amplitude and the central amplitude of the Proca field increases as the solution moves along the curves of Fig.~\ref{figEFTomega}. The $M-\omega$ ($Q-\omega$) relation for a positive $\lambda$ is presented in~\cite{Minamitsuji:2018kof}. We, however, do not consider the case $\lambda>0$ in this paper because the Proca field with $\lambda>0$ contradicts the standard S-matrix properties such as unitarity and causality~\cite{Adams:2006sv,deRham:2018qqo}.

\begin{figure*}[t]
\centering
 \includegraphics[width=0.8\linewidth]{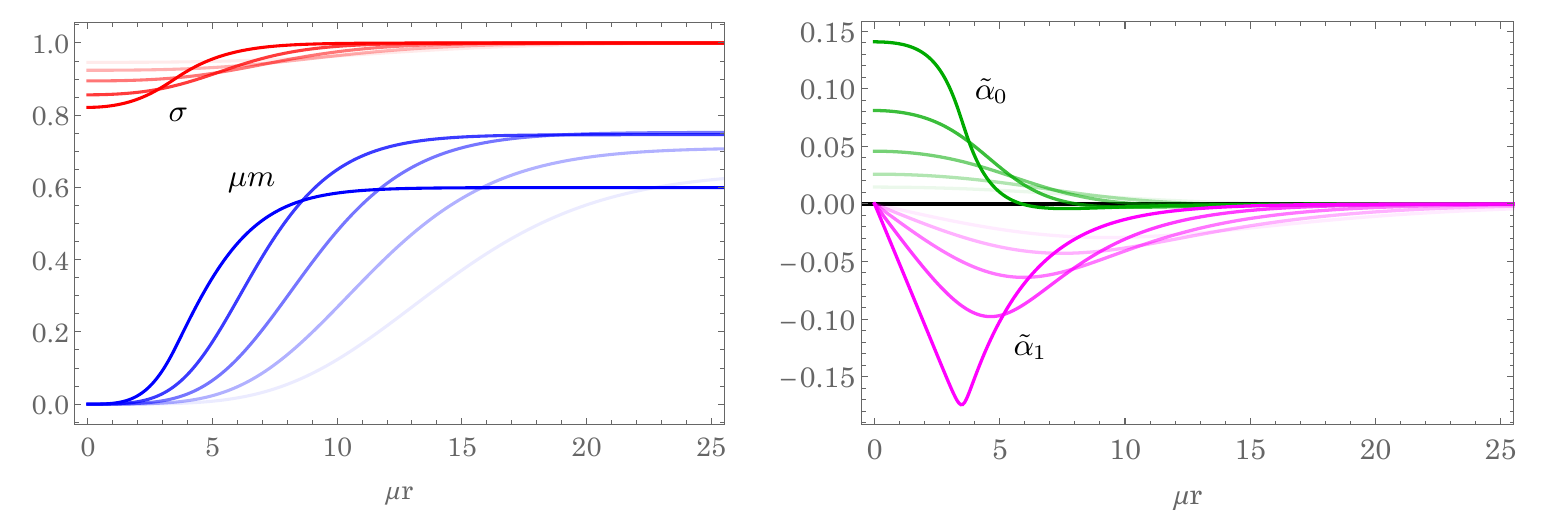}	
\caption{$\sigma$ (red), $\mu m$ (blue), $\tilde{\alpha}_0$ (green), and $\tilde{\alpha}_1$ (magenta) are shown as functions of $\mu r$ with $\tilde{\lambda}=-10$ where a deeper colour corresponds to a larger amplitude of the vector field. We have rescaled $\sigma$ so that $\sigma_{\infty}=1$. }
\label{figEFT}
\end{figure*}

\begin{figure*}[t]
\centering
 \includegraphics[width=\linewidth]{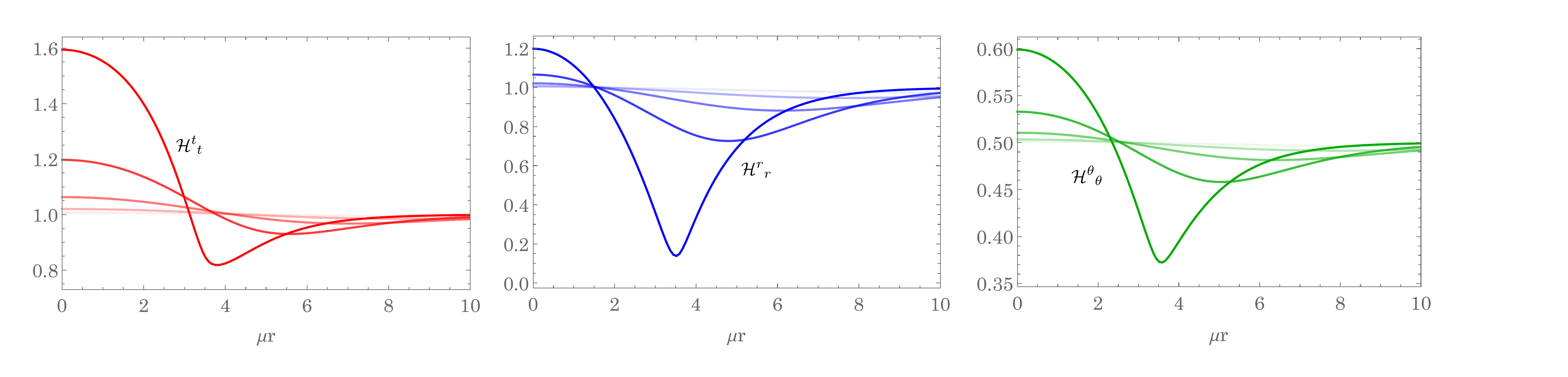}	
\caption{The radial dependence of the components of the effective metric for the solutions presented in Fig.~\ref{figEFT}. }
\label{figEFTmetric}
\end{figure*}

Some numerical solutions and the corresponding effective metric are shown in Figs.~\ref{figEFT} and~\ref{figEFTmetric}, respectively. One may adiabatically
\footnote{
By `adiabatically', we mean that the amplitude of the Proca field is increased with an arbitrary small rate
without breaking the staticity of the background spacetime.
In other words, we consider a family of the static Proca star solutions
with the gradual increase of the central amplitude of the Proca field.}
 increase the amplitude of the Proca field to follow the family of the Proca stars.
As the amplitude of the Proca field increase, the $(rr)$ component of the effective metric approaches zero, forming a singularity at a critical point. The critical point is indicated by a black dot in Fig.~\ref{figEFTomega}. Since the equations of motion are singular at $g_{\rm eff}^{rr}=0$, we cannot find a Proca star beyond the critical point.

In the vicinity of $r=0$,
using \eqref{solution_origin}, ${\cal H}^r{}_r$ can be expanded as 
\begin{eqnarray}
\label{hrr_expansion}
{\cal H}^r{}_r=1-\tilde{\lambda} \tilde{a}_c^2+{\cal O}(r^2),
\end{eqnarray}
where $\tilde{a}_c:=a_c/\Mpl$.
For ${\tilde \lambda}>1/{\tilde a}_c^2>0$, 
${\cal H}^r{}_r$ is initially negative,
but immediately increases as $r$ slightly increases, 
and crosses zero in the vicinity of the center.
Hence, no self-interacting Proca star solutions
exist for ${\tilde \lambda}>1/{\tilde a}_c^2$
because of the onset of a gradient instability around the center.
Moreover, 
in the case $1/{\tilde a}_c^2>\tilde{\lambda}>1/(3{\tilde a}_c^2)$,
although a background Proca star solution exists,
in the vicinity of the center
${\cal H}^t{}_t$ crosses zero and 
the solution would suffer from a ghost instability at the level of linear perturbations.
Here, in the rest we do not consider the case of $\tilde{\lambda}>0$.
On the other hand,
for ${\tilde \lambda}<0$,
from \eqref{hrr_expansion} ${\cal H}^r{}_r>0$ in the vicinity of the center,
however as shown in Fig.~\ref{figEFTmetric},
for a sufficiently large central amplitude
approaches zero
at a radius a few times larger than the Compton radius $\mu^{-1}$.
Since the expansion near the origin \eqref{solution_origin} is no longer valid around such a radius,
for ${\tilde\lambda}<0$
we could not analytically estimate the critical amplitude where ${\cal H}^r{}_r$ vanishes
and a gradient instability appears.
We numerically confirmed 
that even for $|{\tilde \lambda}|={\cal O}(100)$
the features of the $M-\omega$ and $Q-\omega$ relations,
especially the position of the critical points
are qualitatively similar to those in Fig.~\ref{figEFTomega}.
In other words,
for ${\tilde \lambda}<0$,
the appearance of the gradient instability and 
the features of $M-\omega$ and $Q-\omega$ relations
are qualitatively insensitive to the choice of $|{\tilde\lambda}|$.

The recent numerical calculations
\cite{Clough:2022ygm,Coates:2022qia,Mou:2022hqb} have also shown that a singular effective metric is indeed dynamically formed from a generic initial condition. Therefore, a formation of a singular effective metric is not an artificial problem and we need to resolve it in order to discuss a nonlinear regime of the Proca field.

\section{Partial UV completion of self-interacting Proca field}
\label{sec:partialUV}
\subsection{A simple model}
We now seek a partial UV completion of a self-interacting complex Proca field. Here, ``partial'' means that the UV theory is not necessary to be UV complete in the strict sense (e.g.,~renormalizable), but the applicable range of the partially UV complete theory is wider than \eqref{action:Proca}.  We refer to~\cite{Aoki:2021ffc} for a detailed discussion about the partial UV completion in the context of a general k-essence field (see also~\cite{Babichev:2016hys,Babichev:2017lrx,Babichev:2018twg,Mizuno:2019pcm,Mukohyama:2020lsu}). The extension to a complex vector field is straightforward if the potential only involves $X=A_{\mu}\bar{A}^{\mu}$. We consider the following UV completion by means of a heavy field $\chi$:
\begin{align}
S_{\rm UV} = \int \D^4 x \sqrt{-g}\bigg[& \frac{\Mpl^2}{2}R- \frac{1}{4e^2}F_{\mu\nu}\bar{F}^{\mu\nu} - \frac{1}{2}f(\chi) X 
\nn
& -\frac{1}{2}(\partial \chi)^2  - \Vcal (\chi) \bigg]
\label{action:UVProca}
\end{align}
where the functions $f(\chi)$ and $\Vcal(\chi)$ will be chosen to reproduce the potential $V(X)$ by integrating out $\chi$. The action has no nonlinear interaction in terms of the Proca field. If the field $\chi$ is sufficiently heavy, the equation of motion of $\chi$ may be approximated as
\begin{align}
\frac{1}{2}f'(\chi)X +\Vcal'(\chi) \simeq 0
\label{eq:ProcaApprox}
\end{align}
which can be solved by $\chi=\chi_A (X )$ under the condition
\begin{align}
\frac{1}{2}f''(\chi_A)X +\Vcal''(\chi_A) \neq  0
\,.
\end{align}
The action \eqref{action:UVProca} is reduced to \eqref{action:Proca} with the potential
\begin{align}
V(X)=\Vcal(\chi_A) + \frac{1}{2}f(\chi_A) X
\end{align}
by neglecting the kinetic term of $\chi$ (See also \cite{Zhang:2021xxa,Jain:2022nqu}).
 The corrections coming from the finite kinetic term can be included by employing the derivative expansion which we will discuss later.

Using the two real vector fields $A^1_{\mu}$ and $A^2_{\mu}$, we write the complex vector as
\begin{align}
A_{\mu}=A^1_{\mu} + i A^2_{\mu}
\,,
\end{align}
giving
\begin{align}
F_{\mu\nu}\bar{F}^{\mu\nu}=F^a_{\mu\nu}F^{a\mu\nu}
\,, \quad
A_{\mu}\bar{A}^{\mu} =A^a_{\mu}A^{a\mu}
\,,
\end{align}
where $a=1,2$ and the summation of $a$ is understood. We introduce St\"{u}eckelberg fields according to the following replacements
\begin{align}
A^a_{\mu} \to \partial_{\mu}\phi^a 
+e A^a_{\mu}  =: D_{\mu}\phi^a
\,.
\end{align}
The action \eqref{action:UVProca} is then
\begin{align}
S_{\rm UV} = \int \D^4 x \sqrt{-g}\bigg[ &\frac{\Mpl^2}{2}R- \frac{1}{4}F^a_{\mu\nu}F^{a\mu\nu} -\frac{1}{2}(\partial \chi)^2 
\nn
&-\frac{1}{2}f(\chi) D_{\mu}\phi^a D^{\mu}\phi^a -\Vcal(\chi) \bigg]
\,.
\label{action:UVProca2}
\end{align}
In particular, the limit $e\to 0$ yields
\begin{align}
S_{\rm UV} \to \int \D^4 x \sqrt{-g}\bigg[& \frac{\Mpl^2}{2}R- \frac{1}{4}F^a_{\mu\nu}F^{a\mu\nu}
\nn
& -\frac{1}{2}\gamma_{IJ}\partial_{\mu}\Phi^I \partial^{\mu} \Phi^J -\Vcal (\chi) \bigg]
\,, 
\label{action:UVProca_e0}
\end{align}
with
\begin{align}
\Phi^I=(\chi,\phi^1,\phi^2)
\,, \quad
\gamma_{IJ}={\rm diag}[1,f(\chi),f(\chi)]
\,,
\label{field_space_metric}
\end{align}
which is nothing but a nonlinear sigma model with decoupled gauge fields. Therefore, the UV theory has neither a ghost nor a gradient instability so long as the field-space metric $\gamma_{AB}$ is positive definite. The field-space metric is invariant under the translations of $\phi^a$ and the rotation. The translation symmetry is gauged in \eqref{action:UVProca2} with the help of the gauge fields $A^a_{\mu}$\footnote{The gauge fields $A^a_{\mu}$ are thus abelian gauge fields. On the other hand, one may consider a UV completion of a complex vector field by the use of a non-abelian gauge field like the $W$ boson in the electroweak theory. In this case, however, the EFT may have another self-interaction $A_{\mu}A^{\mu}\bar{A}_{\nu}\bar{A}^{\nu}$ as well [and have other gauge field(s)] due to the non-abelian origin.} and \eqref{action:UVProca} is recovered in the unitary gauge.

\subsection{Consistency with UV}

We obtain the following relations:
\begin{align}
\frac{\D \chi_A(X)}{\D X} = - \frac{f'(\chi_A)}{2\mu_{\chi}^2}
\end{align}
and
\begin{align}
\label{eft_rel}
V'(X) =  \frac{1}{2} f(\chi_A) \,, \quad
V''(X) = - \frac{1}{4} \frac{f'{}^2(\chi_A)}{\mu_{\chi}^2} 
\end{align}
where $\chi_A$ in the right-hand-side is understood as the solution $\chi_A=\chi_A(X)$ and we have introduced a quantity
\begin{align}
\mu_{\chi}^2 :=  \frac{1}{2}f''(\chi_A)X +\Vcal''(\chi_A)
\,. \label{chimass}
\end{align}
After integrating out $\chi$ while keeping corrections of the kinetic term $(\partial \chi)^2$, the (tree-level) EFT Lagrangian is given by
\begin{align}
\Lcal_{\rm EFT}= \frac{\Mpl^2}{2}R - \frac{1}{4e^2}F_{\mu\nu}\bar{F}^{\mu\nu} -V(X) + \sum_{n=1}^{\infty} \Lcal_{n}
\end{align}
where $\Lcal_{n}$ represent higher derivative corrections which are suppressed by $\partial^{n}/\mu_{\chi}^{n}$ in comparison with the leading term $V(X)$: for instance, the first higher-derivative correction appears at $n=2$ which takes the form
\begin{align}
\Lcal_2 = 
\frac{V''(X)}{2\mu_{\chi}^2(X)}\partial_{\mu} X \partial^{\mu} X
\,.
\label{derivativecorrection}
\end{align}
The mass scale $\mu_{\chi}^2(X)$ determines the cutoff of the derivative expansion and only a finite number of the operators are relevant as long as $\partial \ll |\mu_{\chi}|$. In particular, $\mu_{\chi}$ agrees with the mass of the field $\chi$ around the background $A_{\mu}=0$. Hence, the heavy field $\chi$ is consistently integrated out around $A_{\mu}=0$ only if $\chi$ is not tachyonic, $\mu_{\chi}^2(0)>0$. This requires $V''(0)<0$ which is translated into $\lambda<0$ in \eqref{quarticpotential}. The sign is consistent with the requirement from the S-matrix~\cite{deRham:2018qqo} as it should be.

We emphasize that the higher derivative correction \eqref{derivativecorrection} changes the propagation of the vector field when $A_{\mu}$ has a non-vanishing background configuration. Since \eqref{derivativecorrection} is a higher derivative operator, the resultant dispersion relation of the perturbations must take a nonlinear form,
\begin{align}
g_{\rm eff}^{\mu\nu}k_{\mu} k_{\nu} + \mathcal{O}(k^4/\mu_{\chi}^2) = 0
\,,
\end{align}
rather than the linear form $g_{\rm eff}^{\mu\nu}k_{\mu} k_{\nu}=0$. Even if one finds an instability in the Proca theory (i.e.,~a wrong sign of the effective metric $g^{\mu\nu}_{\rm eff}$), this instability cannot be extrapolated into a UV regime. Appearance of an instability in an IR regime is ubiquitous in many systems (c.f.,~ the dynamical instability of boson stars in the second branch, $M>\mu Q$).

We should, therefore, distinguish whether an instability exists only in the IR regime or continues even in the UV regime. There are two different branches of the singular effective metric, \eqref{reg:effectivemetric1} and \eqref{reg:effectivemetric2}. The second branch \eqref{reg:effectivemetric2} is not singular in the UV theory, suggesting that the singular point that we have encountered in the construction of the Proca star is not an actual singularity from the UV perspective. The point $g_{\rm eff}^{rr} \propto \mathcal{H}^r{}_r=0$ is simply a point that the leading order approximation of the derivative expansion is not valid due to the absence of the leading gradient term. If the derivative correction \eqref{derivativecorrection} is taken into account, the Proca star may continue to exist even beyond the critical amplitude. A similar prescription is well known in the context of the ghost condensate~\cite{Arkani-Hamed:2003pdi}. On the other hand, the first branch $V'=\frac{1}{2}f=0$ is not only a singularity of the effective metric of the EFT but also an (at least ``coordinate'') singularity of the field-space of the UV theory. The function $f$ has to be positive to avoid the UV ghost and, within the regime of validity of the EFT, the positive sign of $f$ is translated into the positivity of $V'$, the effective mass squared of the vector field. This implies that the vector field cannot be tachyonic to be consistent with our UV completion.

It would be worth mentioning a possibility to have an IR ghost in the EFT. Although a UV ghost signals a fatal instability of a theory, an IR ghost around a nontrivial background may be recast in a standard Jeans-like (or tachyonic) instability and does not necessarily render the theory inconsistent~\cite{Gumrukcuoglu:2016jbh}. One can indeed obtain a ghostly EFT from a healthy UV theory without any self-inconsistency~\cite{Garcia-Saenz:2018vqf,Aoki:2021ffc}. However, as opposed to the gradient term, the sign of the kinetic term may not be flipped by a finite number of higher derivative terms to avoid a strong coupling. A distinction between the IR gradient instability and the IR ghost instability is explained in Sec.~V~A of~\cite{Aoki:2021ffc}.
Hence, although $\mathcal{H}^t{}_t =0$ itself is not necessarily pathological in UV, a local EFT cannot achieve to change the sign of the kinetic term. Since we are interested in the asymptotically flat solutions with $\mathcal{H}^t{}_t \to 1$ as $r \to \infty$, we must find $\mathcal{H}^t{}_t >0$ within the validity of the EFT.

In summary, we may conclude
\begin{itemize}
\item $\mathcal{H}^t{}_t$ {\it cannot} change the sign in $\partial \ll |\mu_{\chi}|$;
\item $\mathcal{H}^r{}_r$ {\it can} change the sign in $\partial \ll |\mu_{\chi}|$; and
\item $\mathcal{H}^{\theta} {}_{\theta} \propto V' $ {\it cannot} change the sign in $\partial \ll |\mu_{\chi}|$
\end{itemize}
under the ansatz \eqref{metricansatz} and \eqref{Procaansatz}. Although all the singularities of the Proca theory are apparently pathological, $\mathcal{H}^r{}_r \leq 0$ may be acceptable from the UV perspective. We will confirm these statements by constructing Proca star solutions in the UV theory. Here, we again stress that the Proca theory without higher derivative corrections~\eqref{action:Proca} is not valid at $\mathcal{H}^r{}_r=0$ due to the absence of the leading gradient term. The validity of the Proca theory and the validity of the EFT (the Proca theory with higher derivative corrections) should be distinguished.

\subsection{Reproducing quartic potential}

As an example, we consider the three-dimensional hyperbolic space as the field-space metric:
\begin{align}
f(\chi)=\mu^2 e^{2\chi/\Lambda } \,.
\label{fchi}
\end{align}
The appropriate potential to recover the quartic self-interaction \eqref{quarticpotential} is given by
\begin{align}
\Vcal =-\frac{\mu^4}{4\lambda} \left( 1 -  e^{2\chi/\Lambda} \right)^2
\,.
\label{Vchi}
\end{align}
The field-space metric is regular and the potential is positive semi-definite for $\lambda <0$; thus, there is no pathological instability in the UV theory with $\lambda <0$.
 Our UV theory is characterized by two scales $\Lambda$ and $-\mu^4/\lambda$ where the former one determines the scalar curvature of the field space to be $\mathcal{R}=-6/\Lambda^2$, while the latter one determines the height of the potential. Note that the dimension of the field-space metric
 $\gamma_{IJ}\D \Phi^I \D \Phi^J$ [see Eqs.~\eqref{action:UVProca_e0} and \eqref{field_space_metric}]
is $[{\rm mass}]^2$ rather than $[{\rm length}]^2$.
All the non-renormalizable interactions are suppressed by $\Lambda$ which may justify the use of \eqref{fchi} and \eqref{Vchi} as a {\it partial} UV completion of the Proca theory in energy scales well below $\Lambda$. 
We assume that $\Lambda$ is much larger than the mass of $\chi$ as well as the typical scales of the Proca star discussed below.
\footnote{For this reason, we shall ignore higher derivative corrections to the UV theory which would be suppressed by $\Lambda$ as well.}

The solution to \eqref{eq:ProcaApprox} is
\begin{align}
e^{2\chi_A/\Lambda}  = 1 + \frac{\lambda}{\mu^2} A_{\mu}\bar{A}^{\mu} = \mathcal{H}^{\theta}{}_{\theta}
\label{sol:ProcaApprox}
\end{align}
and then \eqref{action:Proca} with the potential \eqref{quarticpotential} is obtained by substituting this solution to the action by ignoring the kinetic term of $\chi$. One can notice that the solution is consistent only if
\begin{align}
\mathcal{H}^{\theta}{}_{\theta}=1 + \frac{\lambda}{\mu^2} A_{\mu}\bar{A}^{\mu} > 0
\,,
\end{align}
which is equivalent to the condition $f>0$. Since $f=0$ is a coordinate singularity of the field-space metric, one cannot analyse this point in the present variables.
By using the EFT solution \eqref{sol:ProcaApprox}, the mass scale of the heavy field \eqref{chimass} is 
\begin{align}
\label{muchi}
\mu_{\chi}^2 = - \frac{2\mu^4}{\lambda \Lambda^2}\left( 1 + \frac{\lambda}{\mu^2} A_{\mu}\bar{A}^{\mu} \right)^2
\,.
\end{align}
As we have seen, the typical mass $M$ and the radius $R$ of the relativistic Proca star with $|\tilde{\lambda}|=|\lambda| \Mpl^2/\mu^2 = \mathcal{O}(1)$ are of order of
\begin{align}
GM \sim \mu \,, \quad R \sim \mu^{-1}
\,,
\end{align}
with
\begin{align}
A_{\mu} \sim \Mpl
\end{align}
meaning that all the scales of the relativistic Proca star are determined by $\mu$. Therefore, the typical size of the derivative is approximated by $\mu$ and then the condition $\partial^2 \ll |\mu_{\chi}|^2$ in the relativistic regime is given by
\begin{align}
 1 + \frac{\lambda}{\mu^2} A_{\mu}\bar{A}^{\mu} \gg \frac{\Lambda}{|\tilde{\lambda}|^{1/2} \Mpl}
 \label{EFTvalidity}
\end{align}
in the present model.

\section{Proca star beyond critical point}
\label{sec:uvprocastar}

We study the Proca stars
\footnote{
More precisely, the solutions in the partially UV complete theory \eqref{action:UVProca}
should be called scalar-Proca stars due to the existence of the scalar field $\chi$.
Nevertheless, because of the continuation of the argument from the previous sections,
we simply call them Proca stars.}
based on the simple partial UV completion \eqref{action:UVProca} with the hyperbolic field space \eqref{fchi} and the potential \eqref{Vchi}. We use the same ansatz as \eqref{metricansatz} and \eqref{Procaansatz} with $e=1$ and assume a static configuration of the heavy field $\chi=\chi(r)$.

The equations of motion are
\begin{align}
\frac{\D \sigma(r) }{\D r} &= \Fcal_{\sigma}[\sigma, m, a_0, a_1,\chi,\chi', r] \,, \label{UVeom1} \\
\frac{\D m(r) }{\D r} &= \Fcal_m[\sigma, m, a_0, a_1,\chi,\chi', r]  \,, \\
\frac{\D a_0(r) }{\D r} &= \Fcal_0[\sigma, m, a_0, a_1,\chi,\chi', r]  \,, \\
\frac{\D a_1(r) }{\D r} &= \Fcal_1[\sigma, m, a_0, a_1,\chi,\chi', r]  \,, \\ 
\frac{\D^2 \chi(r)}{\D r^2} &= \Fcal_{\chi}[\sigma, m, a_0, a_1,\chi,\chi', r] \label{UVeom5}
\end{align}
where $\Fcal_y~(y=\sigma,m,0,1,\chi)$ are regular functions (modulo the singularities of the spacetime metric), as expected.

The solution near the origin is found to be
\begin{eqnarray}
m(r)
&=&
\frac{\mu^2e^{\frac{2\chi_c}{\Lambda}}}
       {12\lambda\Mpl^2 \sigma_c^2}
\left[
\lambda a_c^2
-2\mu^2  \sigma_c^2 \sinh^2\left(\frac{\chi_c}{\Lambda}\right)
\right]
r^3
+{\cal O} (r^4),
\nonumber\\
\sigma(r)
&=&
\sigma_c
+
\frac{a_c^2\mu^2e^{\frac{2\chi_c}{\Lambda}}}
        {4\sigma_c \Mpl^2}
r^2
+{\cal O} (r^4),
\nonumber\\
a_0(r)
&=&
a_c
+\frac{a_c}{6}
\left(
e^{\frac{2\chi_c}{\Lambda}}\mu^2
- \frac{ {\hat \omega}^2 }{ \sigma_c^2}
\right)
r^2
+
{\cal O} (r^4),
\nonumber\\
a_1(r)
&=&
-\frac{a_c {\hat \omega}}{3 \sigma_c^2 }r
+{\cal O} (r^3),
\nonumber\\
\chi(r)
&=&
\chi_c
-
\frac{e^{\frac{2\chi_c}{\Lambda}}\mu^2}{6\lambda\Lambda }
\left[
\lambda \frac{a_c^2}{\sigma_c^2}
-
\left(
1-e^{\frac{2\chi_c}{\Lambda}}
\right)
\mu^2 
\right]
r^2
+
{\cal O} (r^4),
\nonumber\\
\end{eqnarray}
where $\chi_c$ is the central value of the $\chi$ field.
In the limit of 
$
\lambda \frac{a_c^2}{\sigma_c^2}
-
\left(
1-e^{\frac{2\chi_c}{\Lambda}}
\right)
\mu^2 
\approx 0$,
$\chi$ stays almost constant near the origin
and the solution reduces to that in the EFT \eqref{solution_origin}.

In the limit of $r\to \infty$,
assuming the asymptotic flatness of the spacetime
$m\to m_\infty={\rm const}>0$ and $\sigma\to \sigma_\infty={\rm const}>0$,
then the Proca field behaves as 
\begin{eqnarray}
a_0(r)
&\approx &
a_{0\infty} \frac{e^{-\sqrt{\mu^2-\omega^2}r} }{r} ,
\nonumber\\
a_1(r)
&\approx &
\frac{a_{0\infty} \omega }{\sigma_{\infty} \sqrt{\mu^2-\omega^2} }
 \frac{ e^{-\sqrt{\mu^2-\omega^2}r} }{r} ,
\end{eqnarray}
where $a_{0\infty}$ is a constant and the proper frequency for a distant observer $\omega$ 
is defined by \eqref{proper_freq}.
As $r\to\infty$, 
the equation of motion for $\chi$ asymptotically becomes linear in $\chi$
besides
the source term given by a nonlinear combination of $A_\mu$ and ${\bar A}_\mu$,
and the solution for $\chi$ could be approximately written as 
\begin{align}
\label{chi_asympt_sol}
\chi(r) \approx \chi_+ \frac{e^{\mu_{\chi \infty} r} }{r} + \chi_- \frac{-e^{\mu_{\chi \infty} r} }{r} + 
\chi_{A}(r),
\end{align}
where $\chi_{\pm}$ are integration constants for the homogeneous solutions, 
\begin{align}
\mu_{\chi \infty}^2 := \mu_{\chi}^2(0)= - \frac{2\mu^4}{\lambda \Lambda^2}
\end{align}
is the mass of $\chi$ in the asymptotic region,
and 
$\chi_{A} (r)$
represents the solution of \eqref{eq:ProcaApprox}
which also corresponds to the inhomogeneous part of the solution in the asymptotic region.
The asymptotic flatness requires that the growing solution vanishes, $\chi_+ = 0$.\footnote{Recall that we have assumed $\lambda <0 ~ (\mu_{\chi \infty}^2>0)$ namely a non-tachyonic field $\chi$. In the tachyonic case $\lambda > 0 ~ (\mu_{\chi \infty}^2<0)$, the field $\chi$ shows an oscillating behaviour.} 
The characteristic size of the Proca star is given by 
the Compton radius of the Proca field $ \mu^{-1}$,
while $\chi$ may vary with the short length scale
$\mu_{\chi\infty}^{-1} \ll \mu^{-1}$ for $\chi_- \neq 0$.
However, the EFT description is valid only when the length scale of the system is much larger than the length scale of the heavy field. Therefore, to reproduce the Proca star within the partially UV complete theory, we have to impose $\chi_- = 0$, so that $\chi$ does not vary with the length scale $\mu_{\chi\infty}^{-1}$. 
Thus, from Eq.~\eqref{chi_asympt_sol}
the asymptotic solution for $\chi$
is given by the inhomogeneous solution $\chi(r)\approx \chi_{A}(r)$ which is indeed what we have found in \eqref{sol:ProcaApprox}.
Since $A_\mu$ and ${\bar A}_\mu$ vary with the length scale $\mu^{-1}$,
$\chi_A(r)$ which also varies with $\mu^{-1}$ is only the solution valid within the EFT.

\begin{figure}[t]
\centering
 \includegraphics[width=0.9\linewidth]{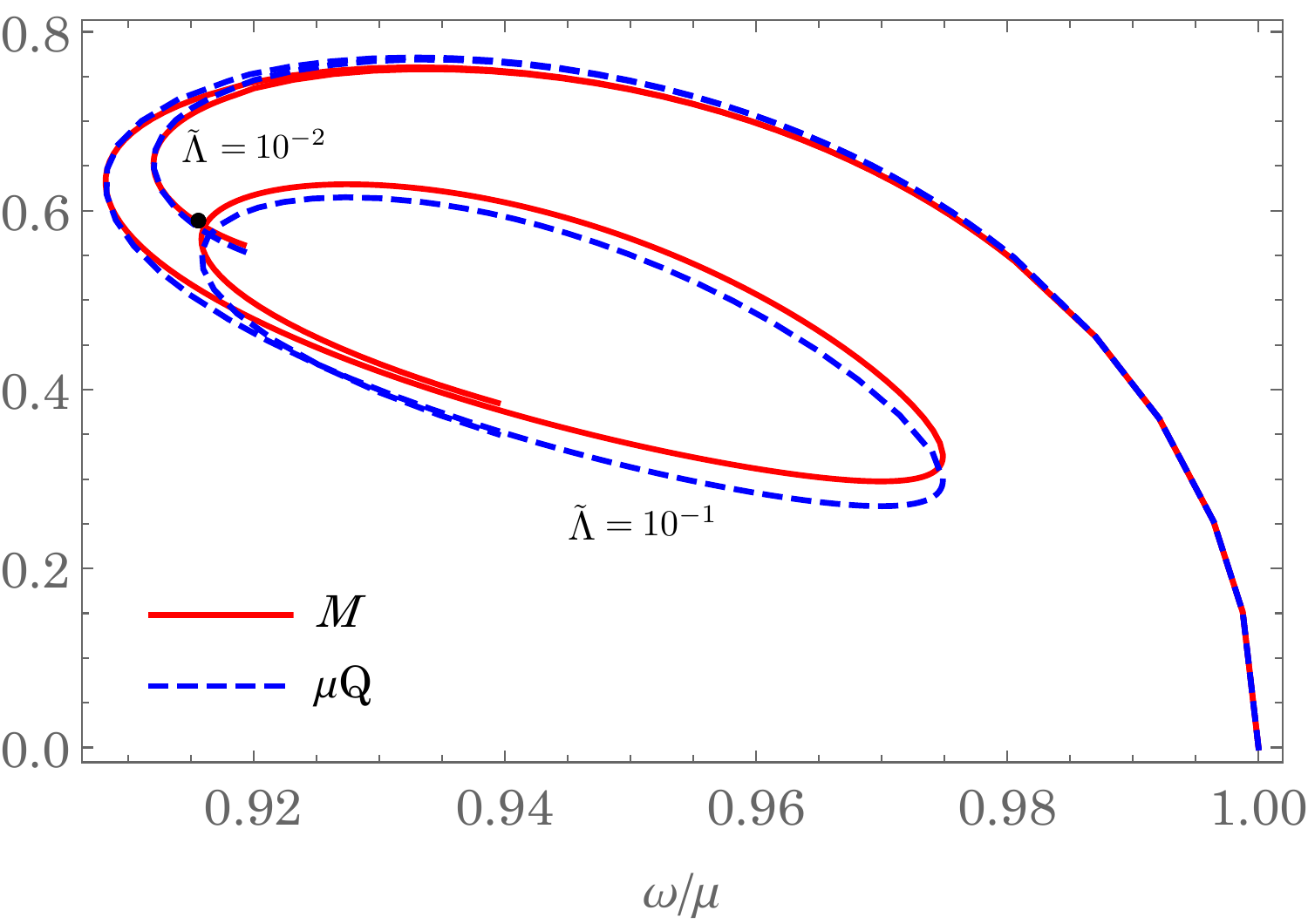}		
\caption{
$M-\omega$ and $Q-\omega$ relations with $\tilde{\Lambda}=10^{-2}$ and $\tilde{\Lambda}=10^{-1}$. }
\label{figUVomega}
\end{figure}

\begin{figure*}[t]
\centering
 \includegraphics[width=\linewidth]{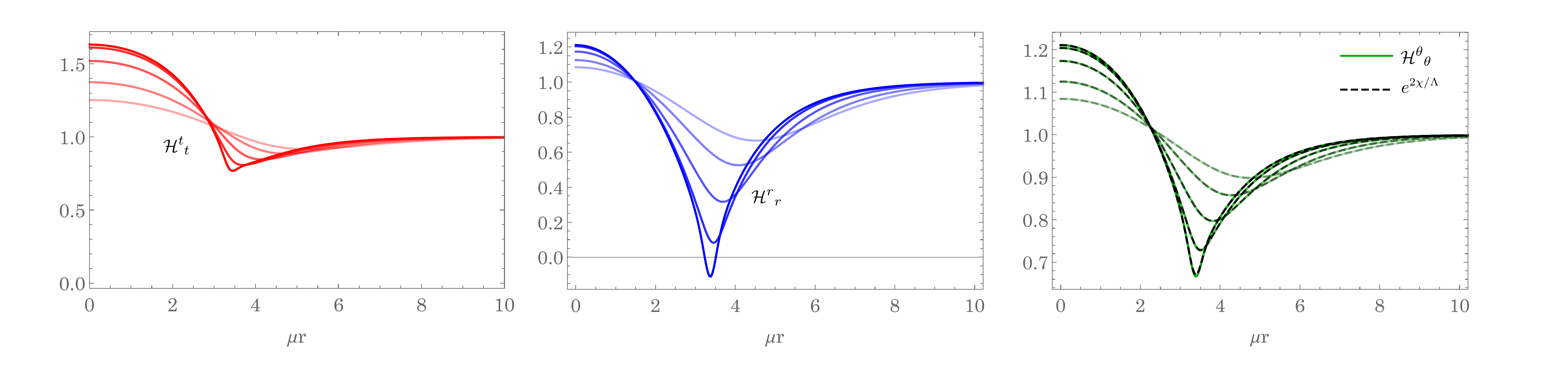}	
\caption{$\mathcal{H}^t{}_t$ (red), $\mathcal{H}^r{}_r$ (blue), $\mathcal{H}^{\theta}{}_{\theta}$ (green), and $e^{2\chi/\Lambda}$ (black dashed) in the partial UV completion of the complex Proca field with $\tilde{\Lambda}=10^{-2}$. 
The thicker curves correspond to the larger central amplitudes of the temporal component
of the Proca field.
}
\label{figUVmetric}
\end{figure*}

\begin{figure}[t]
\centering
 \includegraphics[width=0.8\linewidth]{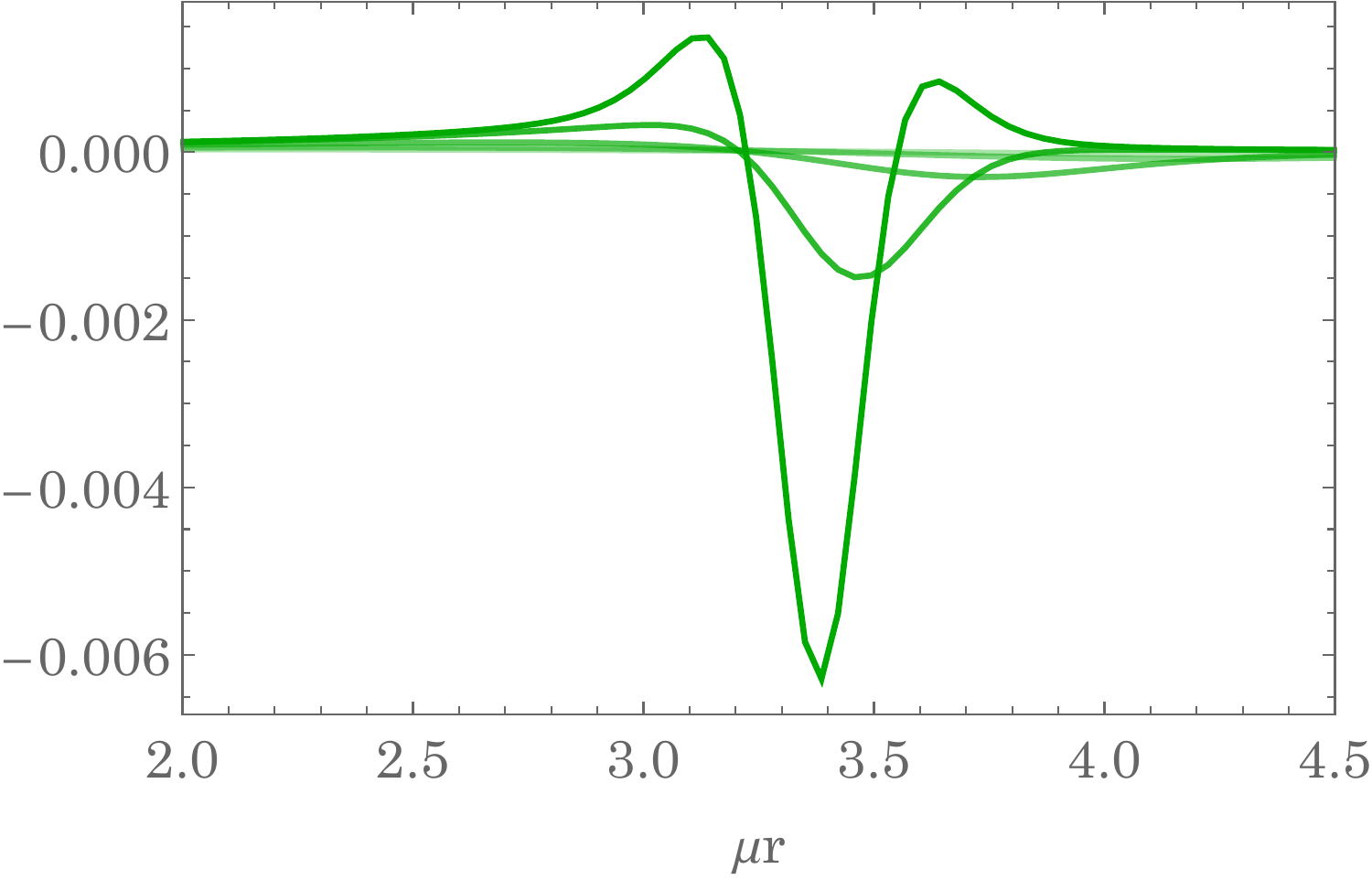}
\caption{
The profiles of $\mathcal{H}^{\theta}{}_{\theta}/e^{2\chi/\Lambda} - 1$ with $\tilde{\Lambda}=10^{-2}$. }
\label{figUVratio}
\end{figure}

\begin{figure*}[t]
\centering
 \includegraphics[width=\linewidth]{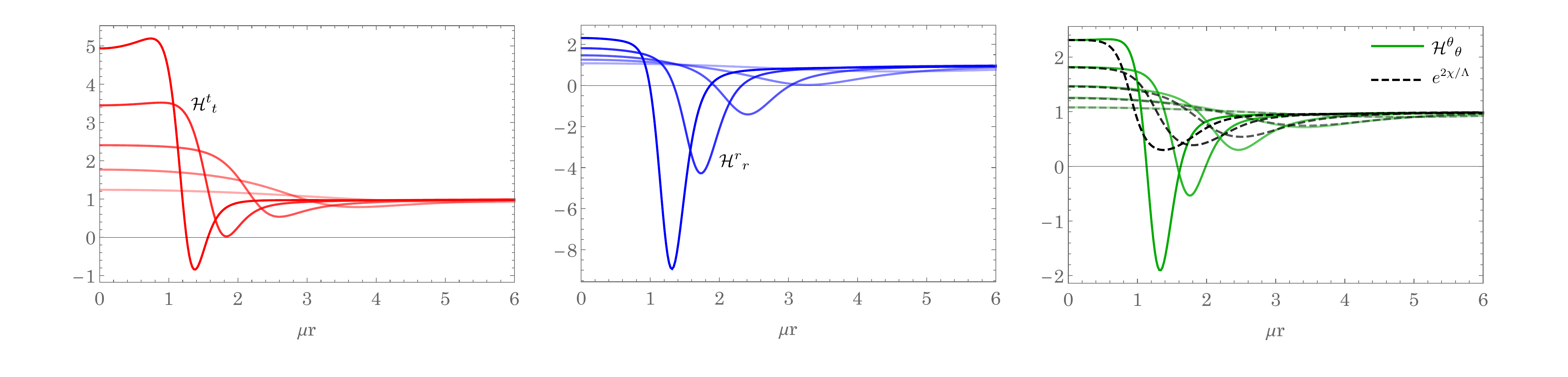}	
\caption{The same figures as Fig.~\ref{figUVmetric} with $\tilde{\Lambda}=10^{-1}$. }
\label{figmildUVmetric}
\end{figure*}

For the numerical calculations, we introduce 
\begin{align}
\tilde{\chi}:=\mu r \frac{\chi}{\Lambda}\,, \quad \tilde{\Lambda}:= \frac{\Lambda}{\Mpl}
\,.
\end{align}
for the $\chi$ field and use the same dimensionless combinations \eqref{normalizevar} and \eqref{normalizepara}. Then, the parameters of the equations of motion are $\tilde{\lambda}$ and $\tilde{\Lambda}$. The boundary condition in terms of $\tilde{\chi}$ is given by
\begin{align}
\begin{cases}
\tilde{\chi} \to 0 \quad {\rm as} \quad \tilde{r} \to 0 \\
\tilde{\chi} \to 0 \quad {\rm as} \quad \tilde{r} \to \infty 
\,.
\end{cases}
\end{align}
As we have discussed, the heavy field $\chi$ should be given by \eqref{sol:ProcaApprox} at the leading-order approximation in the regime of the validity of the EFT. We shall numerically solve \eqref{UVeom1}-\eqref{UVeom5} and use the relation \eqref{sol:ProcaApprox} to check whether solutions in the UV theory agree with the EFT solutions.

The ADM mass $M$ and the Noether charge $Q$ are shown in Fig.~\ref{figUVomega} as functions of $\omega$ for $\tilde{\Lambda}=10^{-2}$ and $\tilde{\Lambda}=10^{-1}$ with $\tilde{\lambda}=-10$. In the case of $\tilde{\Lambda}=10^{-2}$, the result is in good agreement with Fig.~\ref{figEFTomega} up to the critical point indicated by the black dot. Since the equations of motion in the UV theory is regular at $g^{rr}_{\rm eff} \propto \mathcal{H}^r{}_r=0$, we can continue to find the Proca stars beyond the critical point. Fig.~\ref{figUVmetric} represents profiles of the effective metric for some solutions with $\tilde{\Lambda}=1/10$. 
As shown in the right panel of Figs.~\ref{figUVmetric} and \ref{figUVratio}, the profile of $\mathcal{H}^{\theta}{}_{\theta}=e^{2\chi_A(X)/\Lambda}=1 + \frac{\lambda}{\mu^2} A_{\mu}\bar{A}^{\mu}$ almost coincides with that of $e^{2\chi/\Lambda}$. In particular, in the deepest colour curve (the largest amplitude of the vector field), the relation \eqref{sol:ProcaApprox} still holds with an accuracy of one percent though the radial component of the effective metric $\mathcal{H}^r{}_r$ changes the sign, consistently with our discussions in Sec.~\ref{sec:partialUV}: although the point $\mathcal{H}^{r}{}_r=0$ is a singularity of the Proca theory, the EFT description is still available. However, as the amplitude increases, the field configurations become sharper and it is numerically difficult to find a solution. Hence, we also consider a less hierarchical case $\tilde{\Lambda}=1/10$ of which metric profiles are shown in Fig.~\ref{figmildUVmetric}. In this case, we find solutions in which not only $\mathcal{H}^r{}_r$ but also $\mathcal{H}^t{}_t$ and $\mathcal{H}^{\theta}{}_{\theta}$ have zeros. Since $\mu_{\chi}^2$ is proportional to $(\mathcal{H}^{\theta}{}_{\theta})^2$ in the present model, the validity of the EFT description is lost
as $\mathcal{H}^{\theta}{}_{\theta}$ decreases. The right panel of Fig.~\ref{figmildUVmetric} indeed shows that the relation \eqref{sol:ProcaApprox} no longer holds before both $\mathcal{H}^t{}_t$ and $\mathcal{H}^{\theta}{}_{\theta}$ cross zero. Since $g^{\mu\nu}_{\rm eff}$ is the effective metric of the propagation in the regime of the EFT but not the field-space metric of the UV theory, the wrong sign of $g^{\mu\nu}_{\rm eff}$ does not lead to any pathological instability. The positivity of $e^{2\chi/\Lambda}$ guarantees the correct sign of the kinetic term of the UV theory.

Note that the critical point $\mathcal{H}^{r}{}_r=0$ appears at the second branch of the $M-\omega$ relation with $M > \mu Q$. The solution is expected to be dynamically unstable which is consistent with the intuition that the EFT has a gradient instability in the region $\mathcal{H}^{r}{}_r<0$. However, we emphasize that the hyperbolicity of the equations of motion is guaranteed by the positive-definite field-space metric in the partially UV complete theory and there is no pathological instability at UV. The gradient instability is pathological if the instability continues to exist at high momentum modes. On the other hand, by definition, the validity of the EFT is limited in low-momentum modes and the instability in the EFT may not be extrapolated to high-momentum modes. The positive-definite field-space metric in the UV theory implies that the instability exists only in low-momentum modes which is a standard dynamical instability of the high-density boson stars.

\section{Summary}
\label{sec:summary}

It has been argued that a self-interacting Proca field is pathological 
due to a singularity formation in the effective metric 
along which a longitudinal polarization of the Proca field propagates. 
In the numerical construction of Proca star solutions,
such a pathology appears at a critical point where the derivative of the radial Proca profile blows up,
as one cannot numerically integrate the field equations beyond it.
The critical point appeared for a critical amplitude of the Proca field at the center,
beyond which Proca star solutions cease to exist.
We confirmed that at a critical point
the radial component of the effective metric for the self-interacting Proca field vanishes,
and the existence of it may be interpreted as the onset of a gradient instability
at the background level.
The problem is closely related to a ghost instability 
recently claimed in \cite{Clough:2022ygm,Coates:2022qia,Mou:2022hqb},
where  
the time evolution of a self-interacting (real) Proca field crushes 
at a finite time,
and the temporal component of the effective metric for the Proca field vanishes.
The similarity between these two issues indicate 
that they may be solved simultaneously within a single framework
for the extension of the self-interacting Proca theory.
We note that 
a singularity of the effective metric is not a coordinate singularity but a physical one, 
as the conditions \eqref{reg:effectivemetric1} and \eqref{reg:effectivemetric2}
are expressed by coordinate-independent scalar quantities.

We then considered the possibility that
the self-interacting Proca theory is a not fundamental theory, but
a low-energy effective description of a more fundamental theory.
We proposed a simple (partial) ultraviolet (UV) completion model of the self-interacting Proca theory \eqref{action:UVProca}
by introducing the new scalar field $\chi$ 
which is heavy enough in the regime 
where the self-interacting Proca theory arises as the lowest order part of an effective field theory (EFT).
From the EFT viewpoint,
the onset of a gradient or ghost instability may not be a fundamental pathology,
but simply indicates a breakdown of the self-interacting Proca field as an EFT. 
Using this model for the partial UV completion of the self-interacting Proca theory,
we demonstrated that 
Proca star solutions continue to exist even beyond the critical point at which the EFT suffers a gradient instability.
Around a critical point of the EFT, $\mathcal{H}^r{}_r =0$, 
the EFT relation \eqref{sol:ProcaApprox} is slightly violated.
However, a small deviation is enough to regularize the singularity in the EFT,
making Proca star solutions exist beyond it.

By further increasing the amplitude of the Proca field, we found solutions with zeros 
of $\mathcal{H}^t{}_t$
and $\mathcal{H}^{\theta}{}_{\theta}$. In these points, the EFT relation \eqref{sol:ProcaApprox} is violated of the order of unity, meaning that the EFT description is completely broken. One should return to the UV theory if the EFT tends to approach the singularities of $\mathcal{H}^t{}_t = 0$
and $\mathcal{H}^{\theta}{}_{\theta}=0$.

Although we focused on the quartic-order self-interaction \eqref{quarticpotential},
the analysis can be naturally extended to a more general self-interaction
including higher-order terms of $X$.
Note that we focused on the case of $\lambda<0$,
as if  $\lambda>0$ the UV theory \eqref{action:UVProca} becomes pathological
and in the context of the EFT a ghost instability appears
at a lower amplitude than that for a gradient instability.
Moreover, 
we emphasize that the issue on Proca stars
provides one of the simplest cases
to demonstrate the partial UV completion to cure the pathology of the self-interacting Proca field,
in the sense that the system is given by a set of the ordinary differential equations.
On the other hand,
the problems of perturbations of the self-interacting Proca field on a nontrivial background
or the fully nonlinear time evolution of it,
including the cases considered in \cite{Clough:2022ygm,Coates:2022qia,Mou:2022hqb},
are formulated by a set of the partial differential equations
depending on both the space and time.
Thus, the next task should be to confirm 
the well-posedness of initial value problems.
Nevertheless,
since the UV theory \eqref{action:UVProca} was proposed in a general way,
we expect that 
this is generally applied to any type of the breakdown of the self-interacting Proca theory as an EFT.

In particular, a dynamical formulation of the singular effective metric would lead to a spontaneous excitation of heavy degree(s) of freedom. Starting from an initial condition where the EFT is valid and the heavy mode is not excited, the system may evolve into a large amplitude of the vector field as a consequence of a superradiant instability or a gravitational collapse. As performed by~\cite{Clough:2022ygm,Coates:2022qia,Mou:2022hqb}, a singular effective metric dynamically forms in the EFT. However, we have discussed that the formation of the singular effective metric is a sign of the violation of the EFT and the UV physics has to be taken into account before forming it. We may expect that the heavy mode is excited during this dynamical process. If this is indeed the case, these phenomena may be used to extract an observational signature of the underlying theory of the massive vector field.

Finally, we would like to mention the ghost problem in the context of so-called spontaneous vectorization~\cite{Ramazanoglu:2017xbl,Annulli:2019fzq,Kase:2020yhw,Minamitsuji:2020pak},
which describes  a spontaneous growth of Proca hair in a nontrivial strong gravity background
and is analogous to the well-known spontaneous scalarization \cite{Damour:1993hw,Damour:1996ke}.
Recently, it has been shown that 
the instability appearing in the models for spontaneous vectorization
is not a tachyonic instability but a ghost instability \cite{Silva:2021jya,Demirboga:2021nrc} (see also~\cite{Garcia-Saenz:2021uyv}).
By extending the method considered here, one may regard the original models for spontaneous vectorization
with nonminimal matter-Proca coupling
$S=S_g\left[g_{\mu\nu}, \mathcal{A}_\mu\right]+S_m\left[\Omega^2 (Y)g_{\mu\nu}, \Psi_m\right]$ as an EFT,
where $\mathcal{A}_{\mu}$ is a real vector field, $Y:=\mathcal{A}_\mu \mathcal{A}^\mu$,
$\Omega(Y)$ is a regular function of $Y$,
and $S_g$ and $S_m$ are gravitational and matter actions, respectively.
By introducing a scalar field $\chi$ whose dynamics becomes important in the UV regime,
one may consider a UV completion for the models of
spontaneous vectorization as 
$S=S_g
\left[g_{\mu\nu}, \mathcal{A}_\mu,\chi\right]
+S_m\left[\Omega^2 (\chi)g_{\mu\nu}, \Psi_m\right]$.
Since the profile of $\chi$ depends on the environment due to the coupling to the matter field,
 the $\chi$-dependent mass of the vector field can be environment-dependent (this scenario has been investigated in~\cite{Coates:2016ktu}). However, this UV completion does not resolve the problem because the tachyonic instability of the vector field is still a ghost instability even in the UV theory, 
which we observe through a relation analogous to \eqref{eft_rel}. 
As long as considering the Higgs mechanism to provide a mass of the vector field, the mass term is related to the kinetic term which has to be positive definite. We have argued that the EFT should break down before the ghost appears and all predictions of the EFT after appearing of the ghost cannot be trusted. The ghost instability may not be used for the spontaneous growth of the vector field.

\begin{acknowledgments}
The work of K.A. was supported in part by Grants-in-Aid for Scientific Research of the Japan Society for the Promotion of Science, No.~20K14468 and No.~17H06359. 
M.M. was supported by the Portuguese national fund 
through the Funda\c{c}\~{a}o para a Ci\^encia e a Tecnologia (FCT) in the scope of the framework of the Decree-Law 57/2016 
of August 29, changed by Law 57/2017 of July 19,
and the Centro de Astrof\'{\i}sica e Gravita\c c\~ao (CENTRA) through the Project~No.~UIDB/00099/2020.
M.M. also thanks Yukawa Institute for Theoretical Physics for the hospitality
 under the Visitors Program of FY2022.
\end{acknowledgments}

\bibliography{ref}
\bibliographystyle{JHEP}

\appendix

\end{document}